# Deep-ER: Deep Learning ECCENTRIC Reconstruction for fast high-resolution neurometabolic imaging

Paul J. Weiser [a,b,c,*], Georg Langs [c], Wolfgang Bogner [d], Stanislav Motyka [c], Bernhard Strasser [d], Polina Golland [e], Nalini Singh [e], Jorg Dietrich [f], Erik Uhlmann [g], Tracy Batchelor [h], Daniel Cahill [i], Malte Hoffmann [a,b], Antoine Klauser [j,k], Ovidiu C. Andronesi [a,b,**]

[a] Athinoula A. Martinos Center for Biomedical Imaging, Massachusetts General Hospital, Boston, MA, USA
[b] Department of Radiology, Massachusetts General Hospital, Harvard Medical School, Boston, MA, USA
[c] Computational Imaging Research Lab - Department of Biomedical Imaging and Image-guided Therapy, Medical University of Vienna, Vienna, Austria
[d] High Field MR Center - Department of Biomedical Imaging and Image-Guided Therapy, Medical University of Vienna, Vienna, Austria
[e] Computer Science and Artificial Intelligence Lab, MIT, Cambridge, MA, USA
[f] Pappas Center for Neuro-Oncology, Department of Neurology, Massachusetts General Hospital, Boston, MA, USA
[g] Department of Neurology, Beth-Israel Deaconess Medical Center, Boston, MA, USA
[h] Department of Neurology, Brigham and Women's Hospital, Boston, MA, USA
[i] Department of Neurosurgery, Massachusetts General Hospital, Boston, MA, USA
[j] Advanced Clinical Imaging Technology, Siemens Healthineers International AG, Lausanne, Switzerland
[k] Center for Biomedical Imaging (CIBM), Geneva, Switzerland

## ARTICLE INFO

## ABSTRACT

**Introduction:** Altered neurometabolism is an important pathological mechanism in many neurological diseases and brain cancer, which can be mapped non-invasively by Magnetic Resonance Spectroscopic Imaging (MRSI). Advanced MRSI using non-cartesian compressed-sense acquisition enables fast high-resolution metabolic imaging but has lengthy reconstruction times that limits throughput and needs expert user interaction. Here, we present a robust and efficient Deep Learning reconstruction embedded in a physical model within an end-to-end automated processing pipeline to obtain high-quality metabolic maps.
**Methods:** Fast high-resolution whole-brain metabolic imaging was performed at 3.4 $mm^3$ isotropic resolution with acquisition times between 4:11–9:21 min:s using ECCENTRIC pulse sequence on a 7T MRI scanner. Data were acquired in a high-resolution phantom and 27 human participants, including 22 healthy volunteers and 5 glioma patients. A deep neural network using recurring interlaced convolutional layers with joint dual-space feature representation was developed for deep learning ECCENTRIC reconstruction (Deep-ER). 21 subjects were used for training and 6 subjects for testing. Deep-ER performance was compared to iterative compressed sensing Total Generalized Variation reconstruction using image and spectral quality metrics.
**Results:** Deep-ER demonstrated 600-fold faster reconstruction than conventional methods, providing improved spatial–spectral quality and metabolite quantification with 12%–45% ($P<0.05$) higher signal-to-noise and 8%–50% ($P<0.05$) smaller Cramer–Rao lower bounds. Metabolic images clearly visualize glioma tumor heterogeneity and boundary. Deep-ER generalizes reliably to unseen data.
**Conclusion:** Deep-ER provides efficient and robust reconstruction for sparse-sampled MRSI. The accelerated acquisition-reconstruction MRSI is compatible with high-throughput imaging workflow. It is expected that such improved performance will facilitate basic and clinical MRSI applications for neuroscience and precision medicine.

*Abbreviations:* MRSI, Magnetic Resonance Spectroscopic Imaging; ECCENTRIC, ECcentric Circle ENcoding TRajectorIes for Compressed sensing; Deep-ER, Deep learning Eccentric Reconstruction; TGV, Total General Variation; TGV-ER, TGV Eccentric Reconstruction

* Corresponding author.
** Correspondence to: Thirteenth Street, Building 149, Suite 2301, Boston, MA 02129, USA.
*E-mail addresses:* pweiser1@mgh.harvard.edu (P.J. Weiser), oandronesi@mgh.harvard.edu (O.C. Andronesi).

## 1. Introduction

Magnetic resonance spectroscopic imaging (MRSI) is unique in its ability to non-invasively probe a detailed profile of brain metabolism (Maudsley et al., 2021; Bogner et al., 2012). Hence, it is a highly valuable imaging modality employed in fundamental neuroscience (Duarte et al., 2012) and clinical neurology (Öz et al., 2014).

In particular, whole-brain MRSI at ultra-high-field provides comprehensive in-vivo assessment of more than ten neurometabolites simultaneously with high-resolution spatial mapping. However, its potential is not fully realized due to limitations in technical performance. Such MRSI data are essentially 4D (or more), encoding 3 spatial dimensions and 1 (or more) spectral dimensions. The need to encode the spectral dimension with high temporal rate (>1 kHz) and the low signal-to-noise ratio (SNR) of metabolites impose demanding requirements on the MRSI acquisition compared to other MRI modalities. As a result, the acquisition of high spatial resolution MRSI requires acceleration techniques for scan times that are clinically feasible.

Substantial acceleration can be achieved by combining spectral–spatial encoding (SSE) with undersampling techniques (Bogner et al., 2021; Lam et al., 2023). While non-Cartesian undersampled SSE schemes reduce the acquisition time of high-resolution (≈3 mm isotropic) whole-brain MRSI from hours to less than 10 min (Klauser et al., 2023; Hingerl et al., 2020; Ho and Lam, 2020), the reconstruction times for such rapidly acquired data are often prohibitive (hours) with classical algorithms. This represents a significant obstacle to the adoption of fast high-resolution MRSI for human imaging in research and clinical applications.

Although deep learning (DL) methods enable near-instant reconstruction of structural compressed-sensing MRI (Knoll et al., 2020; Hammernik et al., 2021), for high-resolution MRSI the spectral–spatial data size and the feature-parameter space that have to be explored pose great challenges to deep learning reconstruction with today's computational hardware. Furthermore, due to the low SNR of the metabolite signal, exceptional fidelity of the DL reconstruction is required along the spectral dimension to avoid noise amplification and spurious peaks for accurate metabolite quantification. Considering that SNR of MRSI is 3–5 orders of magnitude lower than MRI, stability that may be appropriate for MRI reconstructions is not sufficient for MRSI.

Due to these challenges, only few DL MRSI reconstructions have been shown to date (Lam et al., 2019; Weiser et al., 2021; Nassirpour et al., 2018), with implementations that may limit their generalization and practical applicability to: (1) Cartesian k-space data, (2) pipelines with multiple neural networks to reconstruct different regions of the k-space, (3) single-slice MRSI, and (4) use of spectral dimension that makes the reconstruction dependent on the nucleus, pulse sequence and B0 field.

In the present work, we addressed these limitations and extended DL MRSI reconstruction to non-Cartesian undersampled SSE acquisitions, such as the ECCENTRIC pulse sequence (Klauser et al., 2023, 2024).

ECCENTRIC acquires randomized circular k-space trajectories, simultaneously accelerating two spatial dimensions and providing optimal SNR for high spatial resolution at ultra-high field. ECCENTRIC acquisition results in a 4D matrix size of 64×64×31×451 complex-valued data points for each receive channel, requiring 13.7 GB of memory. Considering that DL requires holding a computational graph with intermediate tensors to compute gradients via backpropagation (Rumelhart et al., 1986; Hecht-Nielsen, 1992; Cireşan et al., 2010), processing the full MRSI acquisition in one-shot exceeds the capability of the GPU hardware most research laboratories have access to.

Due to the challenges of k-space undersampling and k-space point holding information about every spatial voxel making previous DL strategies not applicable (Nassirpour et al., 2018), we adopted a strategy aligning MRSI reconstruction with dynamic 4D MRI (Vishnevskiy et al., 2020). Specifically, we reconstructed each k-space volume independently along the MRSI time dimension, reducing the input data size to 31 MB.

The reconstruction of individual timepoints brings several benefits. (1) It enables us to use the water signal as training data, which can be acquired substantially faster by omitting water suppression in the MRSI pulse sequence and shortening the repetition time. (2) Reconstruction of water MRSI can be validated directly against high-resolution structural MRI. This is an advantage compared to validation using metabolic maps that have less clear structural features due to lower SNR, which can confound the informativity of quality metrics. (3) Because each time point is reconstructed separately, the reconstruction is independent of pulse sequence characteristics such as the echo time, repetition time, B0 field, and nucleus. Hence, the reconstruction has the potential to generalize naturally to varied acquisition parameters and data.

Processing MRSI as dynamic time-series MRI requires reconstructing 3-dimensional image volumes along the time dimension with high fidelity despite the substantially varying contrast and SNR. Advances in DL have lead to an array of neural-network approaches, most often focusing on the reconstruction of structural MRI (Wang et al., 2016; Aggarwal et al., 2018; Hammernik et al., 2018; Hyun et al., 2018; Schlemper et al., 2017; Küstner et al., 2019; Shaw et al., 2020; Singh et al., 2024; Yaman et al., 2020, 2021; Johnson et al., 2023; Zhou et al., 2023; Lee et al., 2024; Zhu et al., 2018). While the majority of these approaches operate either solely in image space or k-space, a recent class of methods demonstrated improved performance by jointly extracting features from both spaces (Souza and Frayne, 2019; Wang et al., 2019; Zhou and Zhou, 2020; Wang et al., 2020; Singh et al., 2022; Zhou et al., 2023). A promising example of these methods, Interlacer (Singh et al., 2022), achieves dual-space feature extraction with recurring layers that separately learn convolutional filters in each space, subsequent to a mixing operation that adds features from each space to the other after taking the appropriate Fourier transform. This strategy proved to outperform state-of-the-art networks across a variety of tasks, most importantly reconstruction of undersampled 2D multi-channel MRI.

The reconstruction network was integrated in an efficient end-to-end processing pipeline for whole-brain 1H-MRSI and was evaluated on phantoms, healthy human volunteers and glioma patients.

## 2. Materials and methods

### 2.1. Human subjects

27 subjects were scanned at the Athinoula A. Martinos Center for Biomedical Imaging with informed consent, including 22 healthy volunteers (12M/10F, 21–49 years) and 5 patients with glioma tumors (demographics and 2021 WHO histo-molecular diagnosis (Louis et al., 2021) are listed in Table 1). All procedures were performed in compliance with relevant laws and institutional guidelines with a protocol approved by the ethics committee (Protocol 2013P001195, July 7th 2022).

### 2.2. MRSI acquisition

Whole-brain 1H-FID-MRSI was acquired with the 3D-ECCENTRIC (Fig. 1, top) pulse sequence (Klauser et al., 2023, 2024) on a 7T scanner (MAGNETOM Terra, Siemens Healthcare, Germany) equipped with a 32Rx/1Tx head coil (NovaMedical, USA) using: 0.9 ms echo-time, 27° excitation flip-angle, 275 ms repetition-time, field-of-view 220x220x105 $mm^3$, matrix size 64x64x31, 3.4x3.4x3.4 $mm^3$ voxel size. For the ($k_x$, $k_y$) encoding the ECCENTRIC circle radius was set to kmax/8 with spectral bandwidth of 2326 Hz without temporal interleaving. The third dimension $k_z$ was encoded by phase encoding with a 3D spherical stack of ECCENTRIC circles partitions. The acquisition was further accelerated (AF=2-4) by random undersampling ECCENTRIC, resulting in acquisition times between 4:11–9:21 min. Low-resolution water calibration data was acquired to map the coil sensitivity profiles for coil combination of metabolic data and to map the $B_0$ field for $B_0$ correction of metabolic data. Water calibration was acquired with the same ECCENTRIC sequence, omitting the water suppression and using a smaller matrix size (22x22x11) and rosette trajectory in 1:16 min. For deep learning training, fully sampled high-resolution water MRSI data were acquired with the same 3D-ECCENTRIC 1H-FID-MRSI sequence, using the same matrix and FOV as the metabolite MRSI, but omitting the water suppression and with a shorter TR=100 ms to reduce the acquisition time to 6:46 min.

**Table 1**
The demographics of glioma patients along with their histological and molecular diagnoses according to the 2021 World Health Organization guidelines: IDH1= isocitrate dehydrogenase 1, 1p/19q codel= codeletion of the short arm of chromosome 1 and the long arm of chromosome 19, ATRX= $\alpha$-thalassemia mental retardation X-linked, MGMT= O6-methylguanine-DNA methyltransferase, TP53= tumor protein p53, C7+/C10-= gain of chromosome 7 and loss of chromosome 10, CDK4= Cyclin-dependent kinase 4, EGFRvIII= Epidermal growth factor variant III.

| Patient | Age/ | Histological diagnosis | | Molecular diagnosis | | |
|---|---|---|---|---|---|---|
| # | Gender | Grade | Type | IDH1 status | 1p/19q codel | Other |
| 1 | 34/F | 3 | Astrocytoma | mutant | not-deleted | |
| 2 | 25/M | 3 | Astrocytoma | mutant | not-deleted | ATRX, TP53 |
| 3 | 66/F | 3 | Oligodendroglioma | mutant | co-deleted | TP53 |
| 4 | 58/M | 2 | Astrocytoma | mutant | not-deleted | |
| 5 | 35/M | 4 | Glioblastoma | wild-type | not-deleted | C7+/C10-, MGMT-, EGFRvIII |

### 2.3. Deep learning MRSI reconstruction

Reconstruction of multi-channel proton MRSI (1H-MRSI) requires multiple, sophisticated steps to extract metabolite signals from the overwhelming background of water and lipid signals in the presence of the inhomogeneous B0 field and produce metabolic maps. We addressed this challenge by implementing an efficient end-to-end 1H-MRSI processing pipeline that integrates physics-based modeling and data-driven machine learning as shown in Fig. 1, including the following specific steps: (1) single channel initialization (2) coil combination, (3) B0 correction, (4) water and lipid removal, (5) image reconstruction, (6) low-rank decomposition, and (7) spectral fitting. The main novelty of the presented pipeline lies in the design of deep neural networks for the removal of nuisance signals (step 4) and image reconstruction (step 5), which is the main topic of this paper. The deep learning reconstruction is described in the following, while details of the other steps are provided in the Supplementary Material.

In this work, we built on and extended the fully convolutional joint-domain Interlacer (Singh et al., 2022) architecture for effective domain transfer to MRSI reconstruction. First, we extended Interlacer to support our specific data requirements by applying image- and k-space convolutions in 3D. We increased the receptive field in the image-space domain using a series of convolutional blocks and performed coil combination with ESPIRiT (Uecker et al., 2014) coil sensitivity maps in each Interlacer layer. Second, we extended the model to non-Cartesian acquisitions via inclusion of an initial gridding layer that samples the non-Cartesian k-space input onto a grid using iNUFT (Bagchi and Mitra, 1996) followed by FFT. Third, we embedded the network and end-to-end optimized it in conjunction with a full MRSI reconstruction pipeline.

The convolutional network (Fig. 1, bottom) takes as input to the first layer the coil-combined undersampled image data ($x_1$) and gridded multi-channel undersampled k-space ($k_1$) to predict fully sampled k-space on a Cartesian grid, which is subsequently transformed to image space, where the loss is computed. In a series of $n_L = 10$ recurrent Interlacer-type layers $f_i$ ($i \in \{1, 2, \ldots, n_L\}$) (Singh et al., 2022), the input $(x_i, k_i)$ to each layer is added back to its output $f_i(x_i, k_i) = (\tilde{x}_i, \tilde{k}_i)$, removing undersampling artifacts by applying incremental corrections to obtain the fully sampled image $x_{FS}$,

$$k_{i+1} = \tilde{k}_i + k_i \quad (1)$$

$$x_{i+1} = \tilde{x}_i + x_i \quad (2)$$

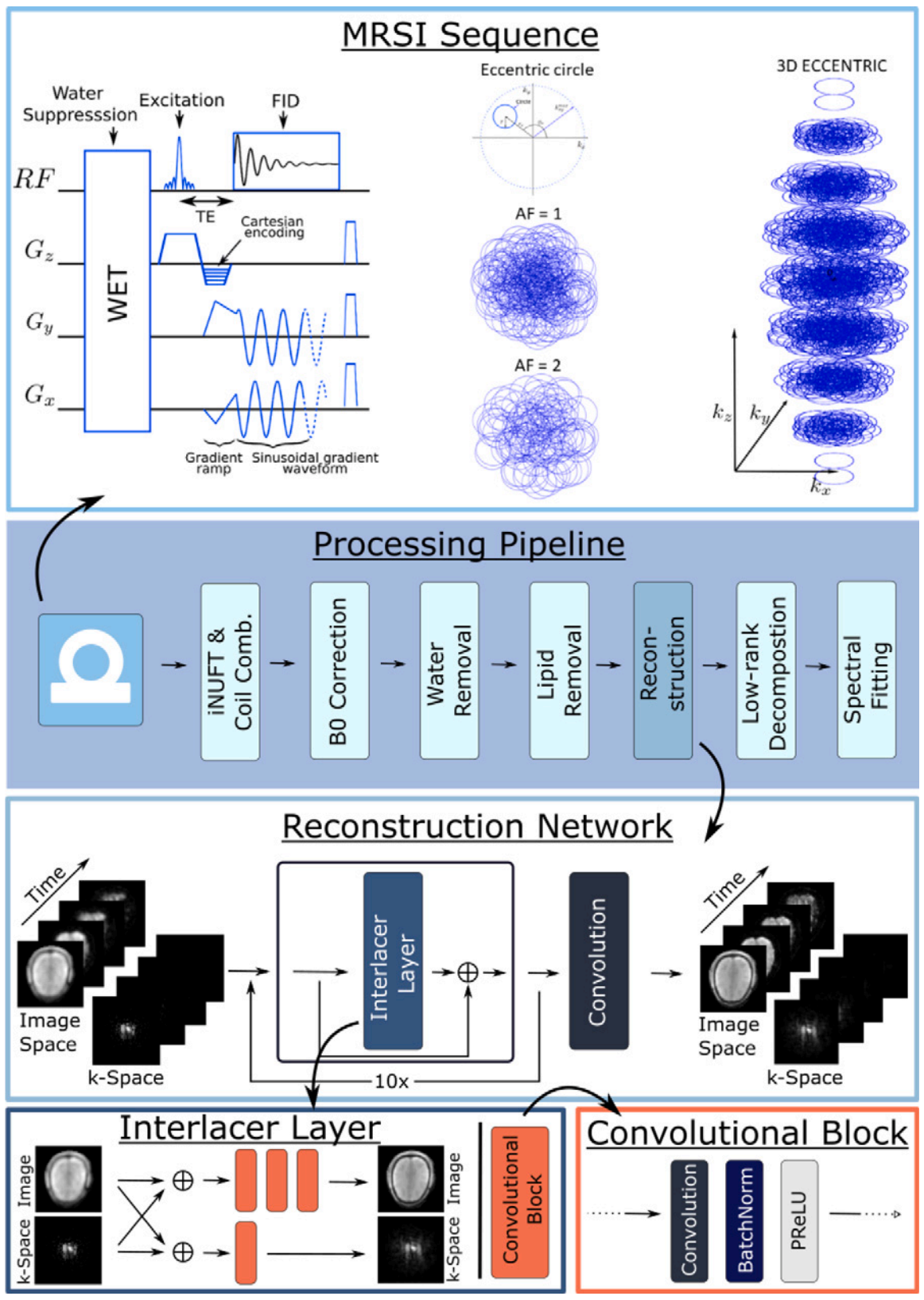

**Fig. 1.** Deep-learning ECCENTRIC reconstruction (Deep-ER), fully compatible with non-Cartesian compressed-sensing MRSI acquisition over the whole brain. Top: ECCENTRIC pulse sequence with ultra-short TE excitation and gradient waveforms for eccentric circles, showing full sampling (AF=1) and twice accelerated compressed-sense undersampling (AF=2), as well as the 3D spherical stack of phase-encoded partitions. Middle: Processing pipeline diagram. Bottom: Deep-learning image reconstruction using 10 fully convolutional Interlacer layers. Each layer processes image and k-space features in parallel, mixing them back together by a learned linear combination after taking the appropriate Fourier transform. The output of each layer is added back to the input. The network reconstructs each 3D time-point of MRSI data separately to maintain independence of the specific acquisition parameters of the MRSI sequence.

Each Interlacer layer separately applies convolutional blocks in k-space and image-space. Before each block, image- and k-space features are merged via weighted addition with learnable mixing parameters $\{\alpha_i, \beta_i\}$,

$$k_i^{mix} = \alpha_i \mathcal{F}(C^{-1}(x_i)) + (1 - \alpha_i)\, k_i \quad (3)$$

$$x_i^{mix} = \beta_i x_i + (1 - \beta_i)\, \mathcal{F}^{-1}(C(k_i)) \quad (4)$$

where $\mathcal{F}$ represents the FFT operation and $C$ the channel-wise coil combination of individual coil images by voxel-wise multiplication with ESPIRiT profiles (Uecker et al., 2014).

A single block with 64 filters is applied in k-space, whereas three blocks with 2-64-2 features, respectively, are applied in image space.

The convolutional blocks each apply 3×3×3 kernels followed by BatchNorm (Ioffe and Szegedy, 2015), as well as ReLU activation in image space and 3-piece activation (Singh et al., 2022) in k-space. Complex values are processed as concatenated real and imaginary channels (Hoffmann et al., 2023b). A final convolutional layer is applied at the end to obtain the desired 64 real and imaginary k-space channels.

The ground truth image $x_{GT}$ was generated for training purposes from fully sampled k-space data $k_{FS}$, utilizing the conventional reconstruction method presented in Klauser et al. (2023, 2024), which is based on an iterative optimization that employs Total-Generalized-Variation (TGV) (Knoll et al., 2011) as a regularizer. Additional details are provided in the TGV-ER subsection of the Methods.

During training, the weights $\theta$ of the neural network $f(\cdot|\theta)$ are optimized subject to

$$\theta = \arg\min_{\theta} \mathbb{E}_{k_{FS}\sim\mathcal{K}}\left[\mathrm{L}\left(f(\mathcal{U}k_{FS}|\theta), x_{GT}\right)\right] \quad (5)$$

where L is a loss function measuring the error of the network prediction from the ground-truth image, $\mathcal{U}$ is the undersampling operator that derives $k_{US}$ from $k_{FS}$, and $\mathcal{K}$ represents the distribution of fully sampled ECCENTRIC k-space training data. The loss function L combines mean-squared-error (MSE) and structural-similarity-index (SSIM) terms,

$$\mathrm{L}\left(f(k_{US}), x_{GT}\right) = \mathrm{MSE}\left(f(k_{US}), x_{GT}\right) + \left(1 - \mathrm{SSIM}\left(f(k_{US}), x_{GT}\right)\right) \quad (6)$$

in order to minimize outliers and maximize visually perceptible structural information between the network output $f(k_{US})$ and the ground truth image $x_{GT}$. In the following, the trained image reconstruction network is referred to as Deep-ER (Deep learning Eccentric Reconstruction).

### 2.4. Deep-ER training details

Water MRSI data from 21 subjects were used for training and validation, with 6 additional subjects used for testing. The training data were augmented by adding a random global phase and random rotation (±0.3 rad), translation (±20 mm), and scaling (±20%) transformations in image space (Hoffmann et al., 2023a).

During training, non-Cartesian k-space data was randomly undersampled to achieve accelerations $AF \in [1, 6]$. Two types of k-space trajectories are acquired by ECCENTRIC: circles that pass through the center of their $k_z$ partition and circles that do not. Retrospective undersampling exclusively omitted the latter, which is in-line with ECCENTRIC undersampling during the acquisition. Data were normalized into the interval $[0, 1]$ in image space by dividing the undersampled input and TGV-ER-reconstructed ground truth data by the maximum absolute value of the input. Optimization used Adam (Kingma and Ba, 2014) with a learning rate of $10^{-5}$ for 500 epochs over the training set. The network was trained on a Dell PowerEdge R7525 server with 64 CPU cores (AMD EPYC 7542 2.90 GHz, 128M Cache, DDR4-3200), 512 GB CPU RAM (RDIMM, 3200MT/s), 3 NVIDIA Ampere A40 GPUs (PCIe, 48 GB RAM) running Rocky Linux release 8.8 (Green Obsidian) using PyTorch 2.2.1 and CUDA 12.1 packages in Python 3.8.

### 2.5. Statistical analysis

Paired one-tail T-Test was used to check statistical significant ($P<0.05$) improvement in a voxel-wise comparison of metabolic image maps obtained by Deep-ER relative to TGV-ER reconstruction. P-values were adjusted for multiple comparison by Bonferroni correction.

**Table 2**
Processing times for TGV-ER and Deep-ER pipelines. 'Image Reconstruction' includes only the time taken by this pipeline's step. The second time includes reconstruction and all the prior steps, while the last line provides the total time that includes also the spectral fitting by LCModel (Provencher, 2014) after the reconstruction. The TGV-ER performs lipid suppression, Fourier transform and B0 correction during the iterative reconstruction.

| | TGV (hh:min.) | Deep-ER (hh:min.) |
|---|---|---|
| Image Reconstruction | 09:50 | 00:01 |
| Pipeline w/o Spectral fitting | 11:23 | 00:28 |
| Pipeline w Spectral fitting | 13:06 | 02:11 |

## 3. Results

To evaluate the performance of the newly developed deep learning Deep-ER reconstruction pipeline we compared its results to the conventional TGV-ER reconstruction pipeline that was previously demonstrated (Klauser et al., 2023, 2024).

In Table 2 the computational efficiency of the Deep-ER and TGV-ER reconstructions are compared. The image reconstruction step of the 4D (k,t) ECCENTRIC data by the Interlaced network (Singh et al., 2022) is performed in 1 min, which is approximately  600 times (590) faster than conventional reconstruction. The total processing times which include all processing steps, with and without spectral fitting, are provided for each MRSI pipeline. Additional speed-up is possible for Deep-ER pipeline due to faster water and lipid removal by the WALINET (Weiser et al., 2025) deep neural network. Hence, in the case of Deep-ER pipeline the largest contribution to the processing time comes from the last step of spectral fitting.

The performance of the Interlacer reconstruction (Deep-ER) was first evaluated on the water MRSI test data acquired in human participants. For this, the water suppression was turned off during ECCENTRIC acquisition while in the processing pipeline the water-lipid removal and spectral fitting steps were omitted. Hence, the quality of the water MRSI data is determined only by the performance of the image reconstruction step. As can be seen by visual inspection of Fig. 2, the water images obtained by Interlacer reconstruction agree well with the ground-truth T1 weighted MRI, showing improved image quality compared to TGV and iNUFT reconstructions. In particular, iNUFT exhibits visible undersampling ringing artifacts for acceleration factors higher than 2. The NRMSE (normalized root mean square error) and SSIM (structure similarity index) show less error and more structural similarity for the Interlacer compared to TGV and iNUFT. While the images above show improvements for the first time point of the FID (free induction decay), the time series FIDs at the bottom indicate that across time dimension the Interlacer provides more stable reconstruction with increasing acceleration. There is higher variability between the FIDs of different accelerations for iNUFT and TGV. At higher accelerations (A.F.$\geq$ 4), there is increased jittering of the FID for iNUFT and TGV reconstructions. Larger FID variability and jittering results in noisier spectra and metabolic maps as can be seen in Supplementary Figure 1.

The MRSI pipeline was evaluated next on the high resolution structural metabolic phantom shown in Fig. 3. The Deep-ER pipeline provides higher quality metabolic images compared to TGV-ER, visualizing well structural features up to 4 mm resolution, which can be resolved by the 3.4 mm resolution of ECCENTRIC acquisition. The correlation coefficients (CC) between the metabolic maps obtained by the two methods show that overall there is a good agreement between the newly proposed DL reconstruction and the conventional established reconstruction. In the case of TGV-ER the 4 mm diameter tubes are blurred and less resolved compared to Deep-ER. The overlaid spectra at the bottom show that Deep-ER provides a more stable spectral reconstruction across accelerations, while TGV-ER shows more spectral variability. Combined these results demonstrate that the network trained only on human brain data sets, generalizes well to very different unseen data sets such as the structural metabolic phantom.

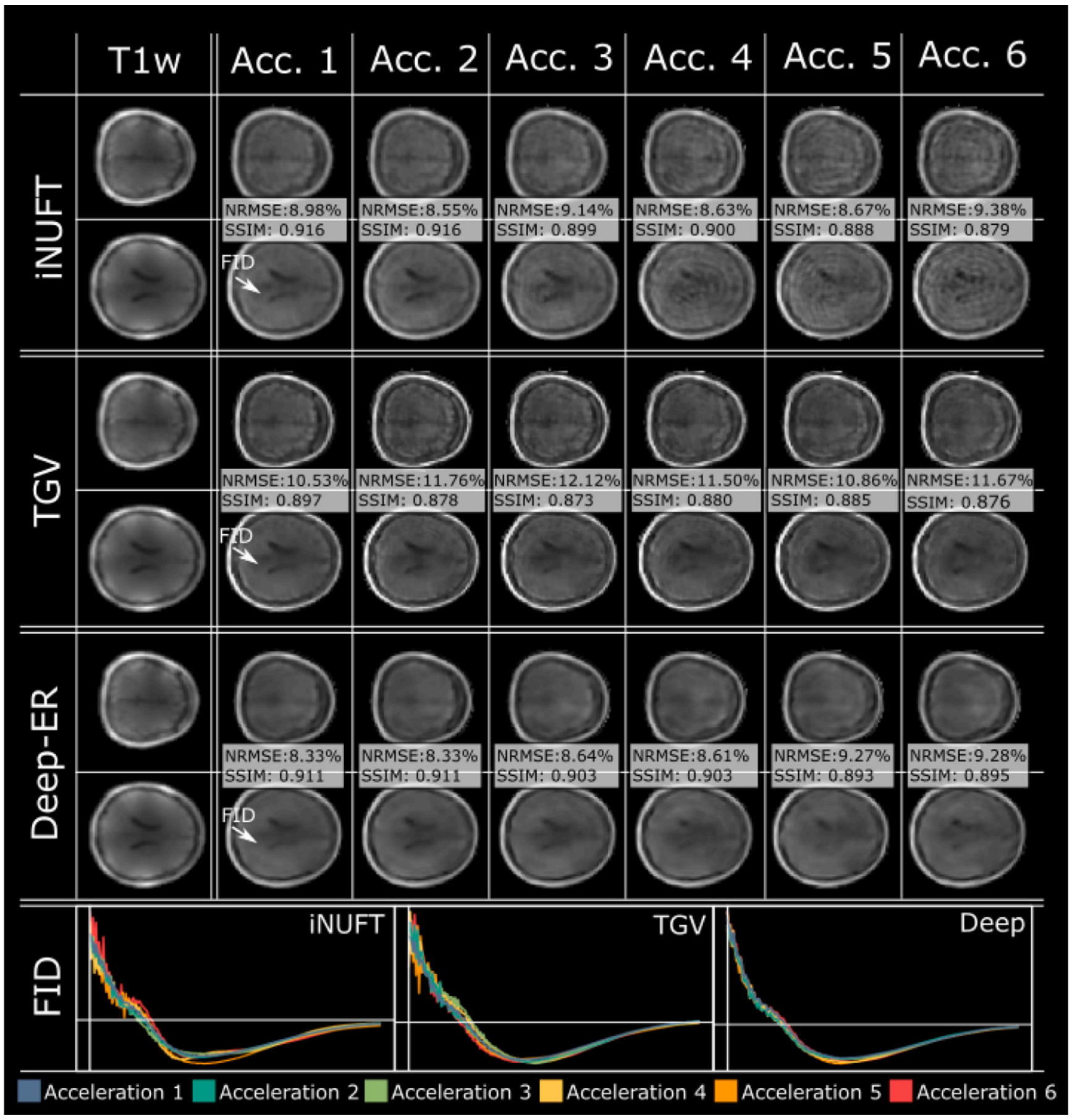

**Fig. 2.** Comparison of reconstruction methods for water images acquired with ECCENTRIC in human brain for accelerations 1 to 6. The top images present the data reconstructed only with the inverse non-uniform FFT (iNUFT). The center images show the reconstruction performed by conventional compressed sense reconstruction (TGV) and the bottom slices show reconstruction by Interlacer (Deep-ER). The images reconstructed for the first FID time point of ECCENTRIC are shown for each reconstruction method. The corresponding ground truth T1-weighted image is shown to the left. Two different slices are presented for each reconstruction method. NRMSE and SSIM were computed for each acceleration between the T1-weighted image and the ECCENTRIC reconstructions. At the bottom examples of FIDs time-series overlaid for all accelerations are shown for all three methods.

In-vivo metabolic images reconstructed from two-fold accelerated (A.F.=2) ECCENTRIC data acquired in a glioma patient and a healthy volunteer are shown in Fig. 4. Metabolic maps in the patient show well defined boundaries for the tumor and metabolic heterogeneity within the tumor. There is higher contrast between the tumor and the normal brain in the maps produced by Deep-ER compared to TGV-ER. In the healthy volunteer similar gray-white matter structural features are visible in the metabolic maps obtained by both Deep-ER and TGV-ER reconstructions. The qualitative parametric maps clearly indicate higher SNR and lower CRLB values for the Deep-ER compared to TGV-ER reconstruction. Examples of spectra show a very distinctive pattern between tumor and healthy metabolic profiles, with a better spectral fit in the case of Deep-ER than TGV-ER spectra. In addition, Supplementary Figure 1 shows metabolic maps obtained for all accelerations (A.F.=1-5) with Deep-ER, TGV-ER and iNUFT reconstructions. For the fully sample data (A.F.=1) the Deep-ER metabolic maps show the sharpest anatomical features compared to TGV-ER and iNUFT maps. As the acceleration increases it can be seen that the metabolic maps obtained by iNUFT reconstruction become gradually noisier, while the metabolic maps of TGV-ER reconstruction exhibit increasing blurring of structural details. In the same time, the accelerated Deep-ER metabolic maps preserve sharper structural features compared to TGV-ER maps and have less noise amplification compared to iNUFT maps. Examples of the spectra show artifacts that overlap metabolite peaks for iNUFT reconstruction of accelerated data, indicative of undersampling aliasing artifacts.

Quantitative analysis from all test subjects is presented in Fig. 5. The Bland-Altman plots show a very small bias between Deep-ER and TGV-ER metabolic maps. The bias increases slightly from the lowest (A.F.=1) acceleration (0.6%–2%) to the highest (A.F.=5) acceleration (3%–7%). A similar trend is noticed for the confidence interval, showing that limits of agreement increases from [−59%, +55%] for A.F.=1 to [−96%, +82%] for A.F.=5. Boxplots indicate Deep-ER has higher SNR (12%–45% more) than TGV-ER, which is statistically significant ($P<0.05$) for reconstructions up to A.F.=4. In addition, Deep-ER has lower CRLB (8%–50% less) than TGV-ER that is statistically significant for A.F.=2. Spectral linewidth with a mean value of 0.04–0.05 ppm is obtained for both reconstructions. Supplementary Figure 2 shows the change in correlation coefficient, normalized root mean square error and structure-similarity index across accelerations. These metrics indicate that as the acceleration increases the difference between Deep-ER than TGV-ER slightly increases (NRMSE from 8% to 12%, SSIM from 0.9 to 0.82, CC from 0.97 to 0.93), however all metrics are well above the thresholds for high agreement and high quality reconstruction.

## 4. Discussion

In this work we developed an end-to-end processing pipeline for high-resolution MRSI data that integrates DL models for image reconstruction and nuisance signal removal. These two steps are time consuming and of critical importance for the quality of metabolic images.

The Deep-ER neural network was specially developed to reconstruct non-cartesian compressed sense MRSI. We demonstrated that Deep-ER provides high efficiency and quality: (1) The reconstruction

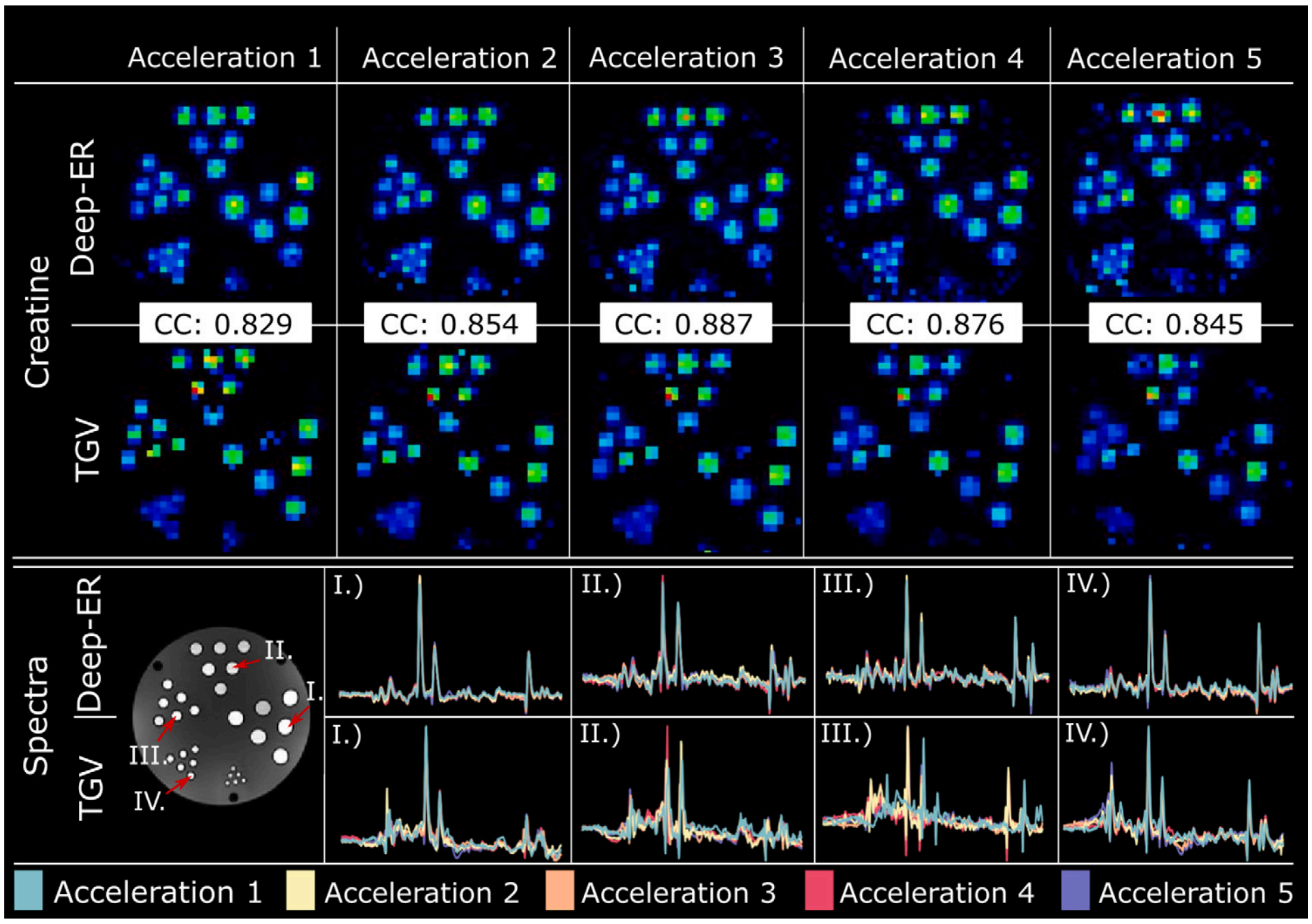

**Fig. 3.** Phantom metabolic images of Creatine reconstructed by Deep-ER and TGV-ER for accelerations from 1 to 5. Correlation coefficients (CC) indicate the agreement between Creatine images reconstructed by the 2 methods. Representative spectra from voxels indicated by arrows are presented at the bottom. Spectra for all accelerations are shown overlaid for each method from the tubes of 10 mm (I), 8 mm (II), 6 mm (III) and 4 mm (IV) diameter. The 2 mm tubes are not individually resolved by the 3.4 mm ECCENTRIC resolution.

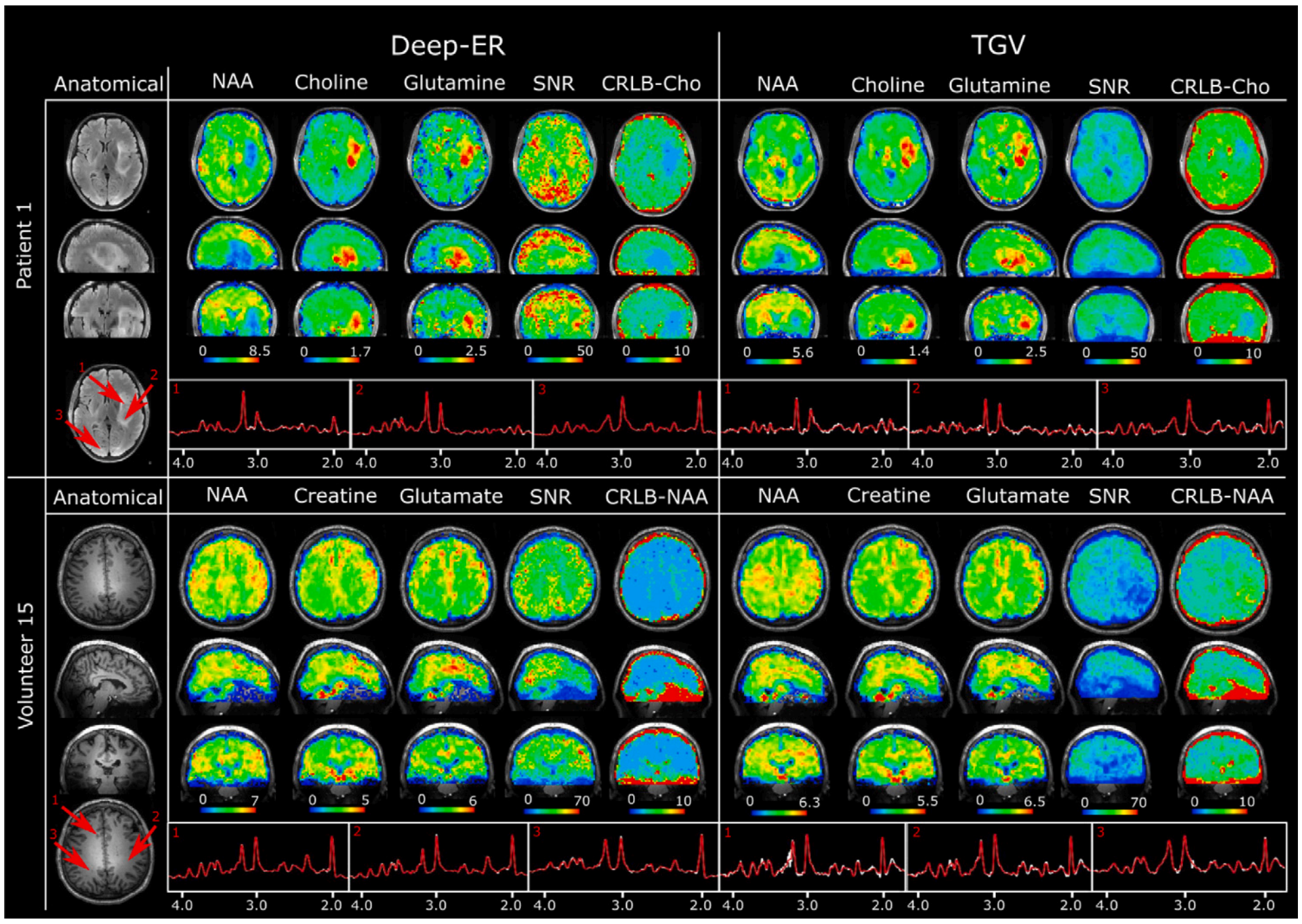

**Fig. 4.** Metabolic images in a glioma patient (top, Patient #1 in Table 1) and a healthy volunteer (bottom). The deep learning Deep-ER reconstruction (left) is compared to the conventional TGV-ER reconstruction (right) showing metabolic maps (NAA, Choline, Creatine, Glutamate and Glutamine), maps of SNR and Cramer–Rao Lower Bounds (NAA and Choline). Example of spectra from individual voxels indicated by red arrows on the anatomical images are shown at the bottom (white trace shows measured spectrum, red trace shows LCModel fit).

time of whole-brain high-resolution 1H-MRSI 3D FID-ECCENTRIC is greatly reduced by a factor of almost 600 using Deep-ER compared to conventional TGV-ER. (2) Efficient usage of GPU memory enables multichannel high-resolution MRSI data processing. (3) High temporal consistency across accelerations reduces spectral noise and improves precision and accuracy of metabolite quantification. (4) Sharper spatial features and less image blurring are achieved with increasing accelerations. Although we demonstrated Deep-ER for undersampled

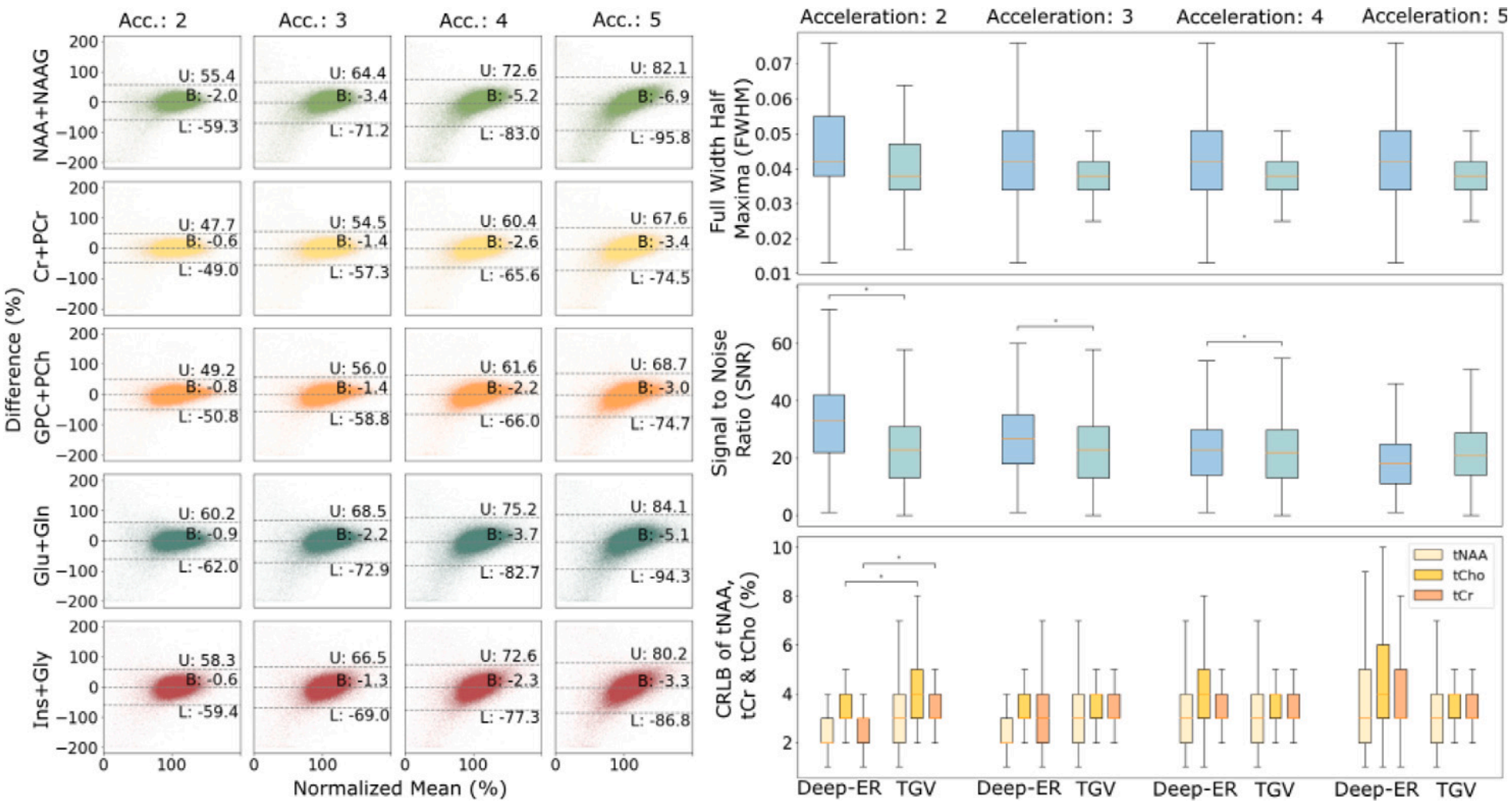

**Fig. 5.** Quantitative comparison of metabolic maps across acceleration factors. Left: Bland-Altman plots are computed for acceleration 2 to 5 for the reconstructed metabolic maps (NAA+NAAG, Cr+PCr, GPC+PCh, Glu+Gln and Ins+Gly). Right: Boxplots of FWHM, SNR and CRLB of total NAA, Choline and Creatine as computed by LCModel. Each plot compares the deep learning based approach Deep-ER to conventional TGV-ER reconstruction across accelerations 2 to 5. Statistical significant differences are indicated by the * symbol.

ECCENTRIC trajectories, the network can be easily applied to other accelerated spectral–spatial encoding trajectories.

Our learning strategy addresses an important bottleneck in neural network development associated with the scarcity of training data for high resolution MRSI. Because of the challenges with acquiring high resolution metabolic MRSI data, it is common practice to use simulated data (Lam et al., 2020) instead of measured data for training models of MRSI reconstruction. However, it is hard to capture by simulations all the complexity of MRSI measurements (Lam et al., 2023). In a departure from prior methods, our approach was to train on the water signal measured by water-unsuppressed MRSI, which is easier to acquire and can provide high quality ground truth images. This concept brings high flexibility and makes the network independent of echo-time, magnetic field and nucleus. Additionally, such training results in a robust network that can generalize the reconstruction to structures that are completely different than brain, as proven on phantom data.

When Deep-ER reconstruction of water-unsuppressed MRSI was compared to high quality T1-weighted MRI we observed very good agreement (NRMSE $\approx$ 8%, SSIM $\approx$ 0.9). The small image difference is explained by small difference in acquisition parameters of water-unsuppressed MRSI by ECCENTRIC and T1-weighted by the standard FLASH sequence (Frahm et al., 1986). Although we tried to match as closely as possible the acquisition parameters (TR, TE, FA, matrix, FOV) of the two sequences, there were few differences such as different RF pulse used for excitation, coil combination and the use of GRAPPA acceleration for T1-weighted. These can lead to slightly different image contrast due to RF transmit and receive inhomogeneity at ultra-high field.

We observed high stability of the Deep-ER reconstruction across the FID time series with increasing acceleration, in particular for the late time points that have significantly lower SNR compared to the beginning of the FID. By comparison there is more variability between FIDs of different accelerations and more FID jittering at high accelerations for TGV-ER and iNUFT reconstructions. The high fidelity of Deep-ER was important to obtain better spectra and correspondingly metabolic maps of superior quality across accelerations, compared to TGV-ER or iNUFT reconstructions.

At the moment the main limitation of Deep-ER is the matrix size of the data, which for this work was fixed to 64x64x31. However, due to its efficient memory utilization and training data generation the network can be trained for higher matrix sizes, and this will be subject of future work. Further acceleration of the end-to-end MRSI processing pipeline can be obtained by speeding with deep learning the spectral fitting (Gurbani et al., 2019; Shamaei et al., 2023a,b; Zhang and Shen, 2023; Chen et al., 2024) and pre-processing steps (coil combination (Motyka et al., 2021), B0 correction (Motyka et al., 2024) or artifact removal/denoising (Rakić et al., 2024; Wang et al., 2023)). In addition, the MRSI data quality can be further improved using real-time motion correction and shim update (Bogner et al., 2013) in combination with integrated receive-shim arrays (Esmaeili et al., 2020).

A particular challenge for the reconstruction of undersampled whole-brain 1H-MRSI data is represented by the overwhelming nuisance signals of water and fat, which create aliasing artifacts during image reconstruction. Nuisance signals need to be removed prior to metabolic image reconstruction, however, conventional methods (Cabanes et al., 2001; Bilgic et al., 2014) come with significant processing times. For this reason we developed the WALINET neural network (Weiser et al., 2025) that removes efficiently and accurately the fat and water signals before reconstruction of metabolite images.

In summary, we demonstrate a robust and fast MRSI processing pipeline that can be combined with accelerated high-resolution MRSI acquisition to obtain high quality metabolic imaging of the brain. We expect that such advanced joint acquisition-reconstruction MRSI methodology will open new avenues of discovery in neuroscience research and enable high-throughput workflow consistent with the needs of clinical translation for precision medicine in patients.

## CRediT authorship contribution statement

**Paul J. Weiser:** Writing – review & editing, Writing – original draft, Visualization, Validation, Methodology, Investigation, Data curation. **Georg Langs:** Supervision. **Wolfgang Bogner:** Supervision. **Stanislav Motyka:** Methodology. **Bernhard Strasser:** Methodology. **Polina Golland:** Supervision, Methodology. **Nalini Singh:** Methodology. **Jorg Dietrich:** Data curation. **Erik Uhlmann:** Data curation. **Tracy Batchelor:** Data curation. **Daniel Cahill:** Data curation. **Malte Hoffmann:** Writing – review & editing, Writing – original draft, Supervision, Software, Methodology, Investigation, Conceptualization. **Antoine Klauser:** Writing – review & editing, Writing – original draft, Software, Methodology, Data curation, Conceptualization. **Ovidiu C. Andronesi:** Writing – review & editing, Writing – original draft, Visualization, Validation, Supervision, Methodology, Investigation, Funding acquisition, Data curation, Conceptualization.

## Funding

NIH, United States grants 2R01CA211080-06A1, R01CA255479, U24NS129893 and P50CA165962 and Austrian Science Fund: WEAVE I 6037 & P34198.

## Declaration of competing interest

The authors declare that they have no known competing financial interests or personal relationships that could have appeared to influence the work reported in this paper.

## Appendix A. Supplementary data

Supplementary material related to this article can be found online at https://doi.org/10.1016/j.neuroimage.2025.121045.

## Data and code availability statement

Data are available from authors based on reasonable request and with signed data shared agreement between institutions. Code is available upon publication on Github: https://github.com/weiserjpaul/Deep-ER.